\documentclass[aps,prl,twocolumn,groupedaddress]{revtex4} 

\newcounter{col}
\usepackage[pdftex]{graphicx}
\usepackage{rotating}
\usepackage{subfigure}
\usepackage{verbatim}
\usepackage{amsmath}
\usepackage{amssymb}
\usepackage{color}
\usepackage{ifthen}

\newcommand{\beq}{\begin{equation}}
\newcommand{\eeq}{\end{equation}}
\newcommand{\beqn}{\begin{eqnarray}}
\newcommand{\eeqn}{\end{eqnarray}}
\newcommand{\avg}[1]{\langle{#1}\rangle}
\newcommand{\ket}[1]{|{#1}\rangle}
\newcommand{\bra}[1]{\langle{#1}|}
\newcommand{\ip}[2]{\langle{#1}|{#2}\rangle}
\renewcommand{\H}{\hat{H}}

\begin{document}

\title{Statistical properties of multistep enzyme-mediated reactions}

\author{Wiet H. de Ronde\footnote{These authors contributed equally to this work}}
\email{deronde@amolf.nl}
\affiliation{FOM Institute for Atomic and Molecular Physics, Kruislaan 407, 1098 SJ, Amsterdam}

\author{Bryan C. Daniels\footnotemark[1]}
\email{bcd27@cornell.edu}
\affiliation{Laboratory of Atomic and Solid State Physics, Cornell University, Ithaca, NY 14853, USA}

\author{Andrew Mugler\footnotemark[1]}
\email{ajm2121@columbia.edu}
\affiliation{Department of Physics, Columbia University, New York, NY 10027, USA}

\author{Nikolai A. Sinitsyn}
\email{nsinitsyn@lanl.gov}
\affiliation{Computer, Computational and Statistical Sciences Division, Center for Nonlinear Studies, Los Alamos National Laboratory, Los Alamos, NM 87545, USA}

\author{Ilya Nemenman}
\email{nemenman@lanl.gov}
\affiliation{Computer, Computational and Statistical Sciences Division, Center for Nonlinear Studies, Los Alamos National Laboratory, Los Alamos, NM 87545, USA}

\date{\today}

\ifthenelse{\value{col} = 1}{\linespread{1}}{}
\begin{abstract}
  Enzyme-mediated reactions may proceed through multiple intermediate
  conformational states before creating a final product molecule, and
  one often wishes to identify such intermediate structures from
  observations of the product creation. In this paper, we address this
  problem by solving the chemical master equations for various
  enzymatic reactions. We devise a perturbation theory analogous to
  that used in quantum mechanics that allows us to determine the first
  ($\avg{n}$) and the second ($\sigma^2$) cumulants of the
  distribution of created product molecules as a function of the
  substrate concentration and the kinetic rates of the intermediate
  processes. The mean product flux $V=d\avg{n}/dt$ (or
  ``dose-response'' curve) and the Fano factor $F=\sigma^2/\avg{n}$
  are both realistically measurable quantities, and while the mean
  flux can often appear the same for different reaction types, the
  Fano factor can be quite different. This suggests both qualitative
  and quantitative ways to discriminate between different reaction
  schemes, and we explore this possibility in the context of four
  sample multistep enzymatic reactions. We argue that measuring both
  the mean flux and the Fano factor can not only discriminate between
  reaction types, but can also provide some detailed information about
  the internal, unobserved kinetic rates, and this can be done without
  measuring single-molecule transition events.
\end{abstract}
\ifthenelse{\value{col} = 1}{\linespread{1.5}}{}

\maketitle

\begin{figure}
\begin{tabular}{|c|c|} \hline
\ifthenelse{\value{col} = 1}{
{\LARGE A} & \scalebox{.5}{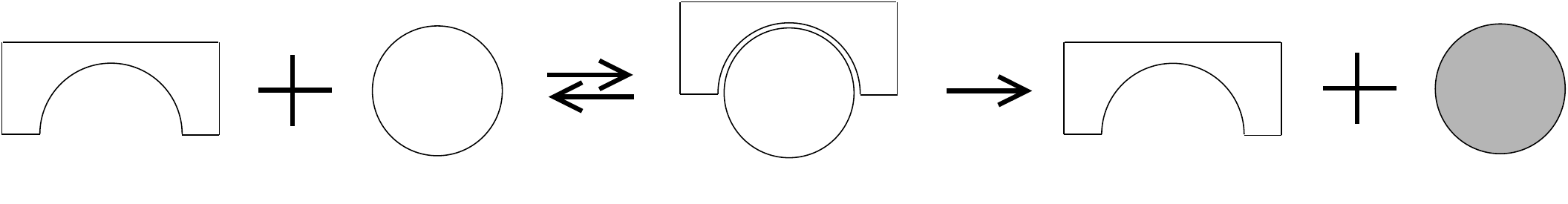}\\ \hline
{\LARGE B} & \scalebox{.5}{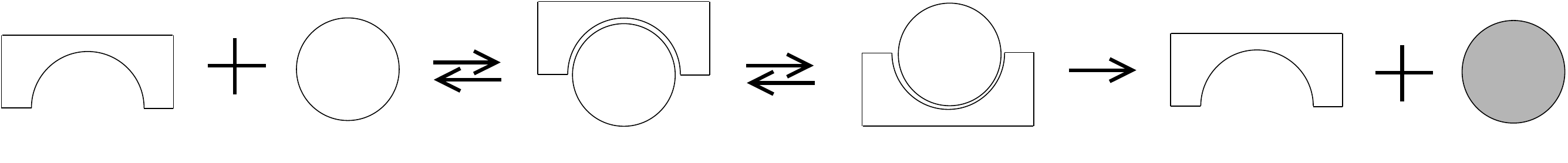}\\ \hline
{\LARGE C} & \scalebox{.5}{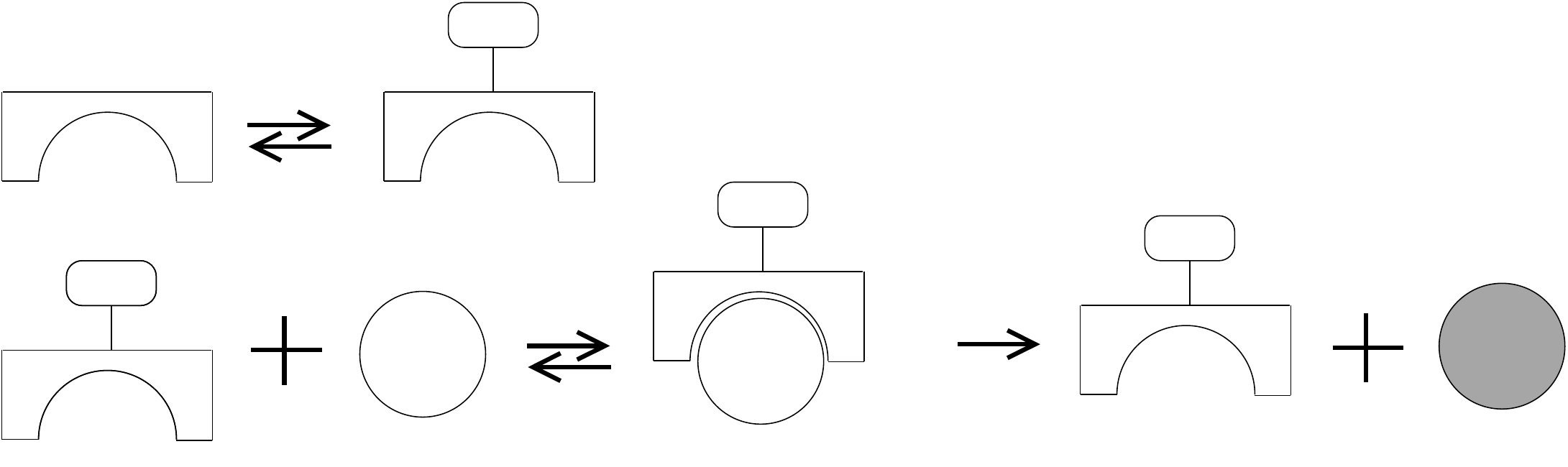}\\ \hline
{\LARGE D} & \scalebox{.5}{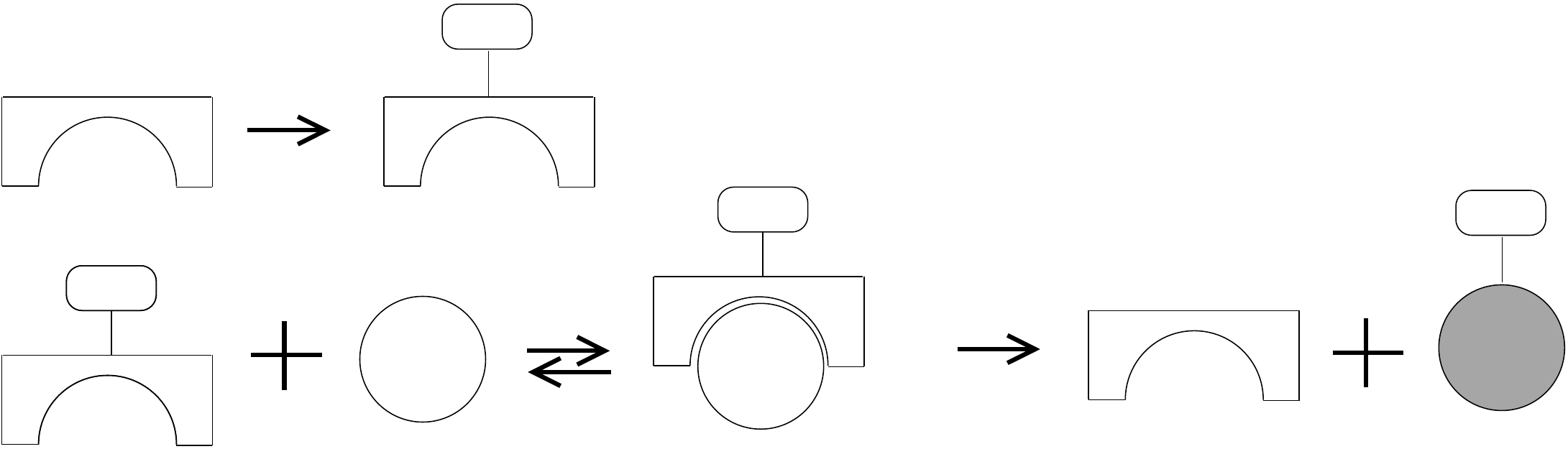}\\ \hline
}{
{\LARGE A} & \scalebox{.28}{\input{cartoon_A.pdftex_t}}\\ \hline
{\LARGE B} & \scalebox{.28}{\input{cartoon_B.pdftex_t}}\\ \hline
{\LARGE C} & \scalebox{.28}{\input{cartoon_C.pdftex_t}}\\ \hline
{\LARGE D} & \scalebox{.28}{\input{cartoon_D.pdftex_t}}\\ \hline
}
\end{tabular}
\linespread{1}
\caption{Potential schemes for an enzyme-mediated reaction, in which
  substrate $S$ is converted to product $P$.  {\bf A:} A simple
  Michaelis-Menten (MM) reaction.  {\bf B:} A MM reaction with an
  additional intermediate state (e.g.\ if the complex undergoes a
  conformational change before creating the product).  {\bf C:} A
  scheme in which the enzyme must become active (e.g., through
  phosphorylation) before mediating the reaction.  {\bf D:} A scheme
  in which the enzyme must become active before mediating the
  reaction, and the reaction leaves the enzyme inactive.}
\label{cartoon}
\end{figure}

Enzyme-mediated reactions are ubiquitous in biology. Traditionally,
they have been described as a two-step Michaelis-Menten (MM) process
\cite{Michaelis}, in which the enzyme and the substrate form a complex
that can decay either back into the enzyme and the substrate, or
forward into the enzyme and the product (see Fig.~\ref{cartoon}A). The
latter step is usually assumed to be irreversible, leaving three
kinetic rates that specify the reaction. To determine these kinetic
rates, a typical experiment measures the average rate of product
formation (or product ``flux'') $V$ as a function of substrate
concentration $S$ (also called a ``dose-response'' curve), producing a
plot as in Fig.~\ref{plots}A.  Two pieces of information can be
extracted from this plot: the saturating reaction rate $V_{\max}$ and
the Michaelis constant $K$, the substrate concentration at half of the
maximum rate.  Importantly, these two measurements do not specify the
three underlying kinetic rates, thus they do not allow for a full
identification of the reaction processes.

The MM mechanism is not entirely general: many enzyme-mediated
reactions consist of multiple intermediate internal steps (such as
conformational changes of either the enzyme or the substrate, enzymes
that occur in active and inactive states, etc.), each with its own
forward and backward reaction rates. While measurements of
substrate-enzyme complex formation and product releases are possible
even on a single molecule level in enzymatic kinetics \cite{English}
and in ion channel transport \cite{Rostovtseva,Nestorovich},
typical experiments cannot resolve intermediate steps when measuring
only the average reaction rate since they produce qualitatively
similar curves for $V(S)$.  For example, the mean flux through an
arbitrary complex ion channel that holds at most one large transported
molecule at a time is indistinguishable from that through a simple
channel with just two internal states \cite{Bezrukov}.

An interesting problem then is to determine which experimental
measurements could identify the multistep nature of an enzyme-mediated
reaction without requiring measurements at intermediate steps. We
suggest that this is possible by measuring not only the mean rate but
also the variance in the rate of the creation of product
molecules. Modern experiments can clearly perform this task in
different experimental systems \cite{English,Golding}.

Here we present a general perturbative approach for calculating the
cumulants of a product molecule flux for a given enzymatic reaction
scheme. To illustrate the method, we first apply it to the usual MM
reaction (Fig.~\ref{cartoon}A).  In addition to recovering the
well-known result for the mean rate of product formation as a function
of substrate concentration, we derive the dependence on substrate of
the Fano factor, the ratio of the variance in the number of product
molecules to the mean.  Importantly, our approach is extendible, at
least in principle, to an arbitrary enzyme-mediated reaction scheme,
and we demonstrate this by analyzing three more complex reaction
schemes, shown in Fig.~\ref{cartoon}B-D.  In the context of these
reactions, we show that the dependence of the Fano factor on the
substrate concentration can produce qualitatively different results
for different reaction types, allowing one to distinguish them
experimentally.  In addition, we argue that quantitative features of
the Fano factor measurements can constrain the values of the
underlying kinetic rate constants more tightly than the mean rate
measurements alone. Measurements of higher order product formation
cumulants, if experimentally possible, would allow one to constrain
properties of the reaction even more strongly.

\section{Methods: The Michaelis-Menten Model}

Going beyond a simple description of the mean production of a
particular molecule and making predictions about the intrinsic noise
requires a stochastic description, such as the chemical master
equation (CME) \cite{vanKampen}.  The CME describes the evolution in
time of the joint probability distribution for the copy numbers of all
species involved in a reaction scheme.  For the enzyme-mediated
reactions we consider, we make the assumption that each enzyme acts
independently, that is, the substrate concentration is much larger
than the enzyme concentration. This is equivalent to treating the
process as if only one enzyme were present at a time.  Furthermore, we
assume that the concentration of the substrate is constant during each
experimental measurement, and thus our master equation needs only to
keep track of the enzyme's state and the number of created product
molecules $n$. We note that both of these assumptions can be relaxed
using recently developed techniques
\cite{Sinitsyn,Sinitsyn2}. Finally, we only search for the
distribution of the number of product molecules at times much longer
than a typical enzymatic turnover time.

We begin by demonstrating our method on the simple Michaelis-Menten
(MM) reaction in Fig.\ \ref{cartoon}A.  In the MM reaction, the enzyme
will be in either a free state $E$ or a bound state $ES$.  Therefore
we partition the joint probability distribution into two parts:
$P^E_n$, the probability that $n$ product molecules have been created
{\it and} the enzyme is free, and $P^{ES}_n$, the probability that $n$
product molecules have been created {\it and} the enzyme is bound,
yielding the CME \cite{vanKampen} \beqn
\label{ma1}
\frac{dP^E_n}{dt}&=&-k_1SP^E_n+k_{-1}P^{ES}_n+k_2P^{ES}_{n-1}\\
\label{ma2}
\frac{dP^{ES}_n}{dt}&=&k_1SP^E_n-(k_{-1}+k_2)P^{ES}_n \eeqn where the
rates are defined in Fig.~\ref{cartoon}A, and $S$ is the number of
substrate molecules. (Note that $S$ can equivalently be thought of as
the concentration of substrate as long as one appropriately rescales
the rates).  The total probability of having $n$ product molecules is
then $P_n=P^E_n+P^{ES}_n$.

We note that the situation where the product molecules are created and
never destroyed or transformed back into the substrate is not
physical, and additional reactions that degrade the product in some
way are needed. However, as long as we are interested in how many
product molecules have been created, rather than are present at a
given time, the creation, Eqn.~(\ref{ma1}, \ref{ma2}), and the decay
reactions can be considered independently.

Similar to Refs.~\cite{Bagrets,Sinitsyn,Sinitsyn2,Gopich,Hornos} and
others, we begin our solution of Eqns.\ (\ref{ma1}-\ref{ma2}) by
defining the generating function \beq G^z(\chi) = \sum_{n=0}^{\infty}
P^z_n e^{i\chi n} \eeq with $z \in \{E, ES\}$.  Defining the vector
$\ket{G}=(G^E,G^{ES})^T$, we may write the total generating function
as \beq G(\chi) = \ip{\hat{1}}{G} = G^E+G^{ES} \eeq where
$\bra{\hat{1}}=(1,1)$ (note that we are adopting ``bra-ket'' vector
notation commonly used in quantum mechanics literature).  The
advantage of this formalism is that the mean $\avg{n}$ and variance
$\sigma^2$ of the distribution of product molecules $P_n$ can be
calculated from $G(\chi)$ via \beq
\label{cu}
\avg{n} = \left.\frac{d(\ln G)}{d(i\chi)}\right|_{\chi=0}, \qquad
\sigma^2 = \left.\frac{d^2(\ln G)}{d(i\chi)^2}\right|_{\chi=0}.
\eeq
Furthermore we note that having $N$ (independently acting) enzymes is equivalent to taking $G$ to $G^N$, so that extension to larger concentrations of enzymes is straightforward.

Now multiplying Eqns.\ (\ref{ma1}-\ref{ma2}) by $e^{i\chi n}$ and
summing over $n$ produces 
\beq
\label{eom}
\frac{d\ket{G}}{dt}=\H \ket{G},
\eeq
where, for the MM reaction,
\beq
\H=\H_A=	\begin{pmatrix}
	-k_1 S	& k_{-1} + k_2e^{i\chi}	\\
	k_1 S	& -(k_{-1} + k_2)
	\end{pmatrix}.
\eeq

Eqn.\ (\ref{eom}) is solved by \beq \ket{G(t)} = e^{\H t}\ket{G_0},
\eeq with an initial condition $\ket{G_0}$.  If we write the matrix
exponential in terms of the eigenvalues $\lambda_j$ and eigenvectors
$\ket{u_j}$ of $\H$ as \footnote{Note that since $\H$ is not symmetric,
  the eigenvectors do not satisfy $\ket{u_j}=\bra{u_j}^T$, but rather they
  solve $\H\ket{u_j}=\lambda_j\ket{u_j}$ and $\bra{u_j}\H=\lambda_j\bra{u_j}$,
  respectively.} \beq e^{\H t} = \sum_j e^{\lambda_j t}
\ket{u_j}\bra{u_j}, \eeq then, at $t$ much larger than the typical enzyme
turnover time, $G(\chi)$ becomes \beq G(\chi) = \sum_j e^{\lambda_j t}
\ip{\hat{1}}{u_j}\ip{u_j}{G_0} \approx e^{\lambda_0 t}
\ip{\hat{1}}{u_0}\ip{u_0}{G_0}, \eeq where $\lambda_0$ is the eigenvalue
with the least negative real part.  Taking the
log, we get \beq
\label{lnG}
\ln G(\chi) = \lambda_0 t + \ln\left(\ip{\hat{1}}{u_0}\ip{u_0}{G_0}\right)
\approx \lambda_0 t, \eeq since again, in the long-time limit, the
first term dominates the second (for any bounded $G_0$), and the
initial number of product molecules is forgotten.  Recalling Eqn.\
(\ref{cu}), it is clear now that one only needs to find the
$\chi$-dependence of the least negative eigenvalue $\lambda_0$ of the
matrix $\H_A$ in order to compute the cumulants of the product molecule
distribution.  In fact, writing $\lambda_0$ as a power series, \beq
\lambda_0 = \sum_{m=0}^\infty \lambda_0^{(m)} \frac{(i\chi)^m}{m!},
\eeq it is clear that one only needs to know the coefficients up to
$m=2$ in order to compute the mean and variance of the distribution;
i.e.\ \beqn
\label{p}
\avg{n} &=& \lambda_0^{(1)} t, \\
\label{sig}
\sigma^2 &=& \lambda_0^{(2)} t, \eeqn and higher order terms are
needed for higher cumulants only.  Since Eqn.~(\ref{cu})
takes $\chi\to 0$, this permits a perturbative approach similar to
that used in quantum mechanics \cite{Griffiths}, with $\chi$ treated as a small parameter.

Specifically, we write $\H=\H^{(0)}+\H^{(1)}\sum_{m=1}^\infty(i\chi)^m/m!$ 
where (for the MM case)
\ifthenelse{\value{col} = 1}{
\beq
\label{MMH}
\H_A^{(0)}=	\begin{pmatrix}
	-k_1 S	& k_{-1} + k_2	\\
	k_1 S	& -(k_{-1} + k_2)
	\end{pmatrix},
\qquad
\H_A^{(1)}=	k_2\begin{pmatrix}
	0	& 1\\
	0	& 0
\end{pmatrix}, \eeq
}{
\beqn
\label{MMH}
\H_A^{(0)}&=&	\begin{pmatrix}
	-k_1 S	& k_{-1} + k_2	\\
	k_1 S	& -(k_{-1} + k_2)
	\end{pmatrix},\\
\H_A^{(1)}&=&	k_2\begin{pmatrix}
	0	& 1\\
	0	& 0
\end{pmatrix}, \eeqn
}
and we truncate at $m=2$.  We emphasize that this truncation does not
introduce any further approximation if one is interested only in the
first and second moments of the product molecule distribution.  
The least negative eigenvalue of $\H^{(0)}$ is $\lambda_0^{(0)}=0$
\footnote{More precisely, $\H_0$ is a propensity matrix whose columns
  sum to zero, which means one of its eigenvalues is zero and the rest
  are negative \cite{vanKampen}.}, and the higher order corrections are
given by \cite{Griffiths}
\beqn
\label{l1}
\lambda_0^{(1)} &=& \bra{u_0^{(0)}}\H^{(1)}\ket{u_0^{(0)}},\\
\label{l2}
\lambda_0^{(2)} &=& \lambda_0^{(1)}-2\sum_{j\ne 0}\frac{1}{\lambda_j^{(0)}}|\bra{u_j^{(0)}}\H^{(1)}\ket{u_0^{(0)}}|^2.
\eeqn

Noting Eqns.\ (\ref{p}-\ref{sig}), the rate of product formation
$V=d\avg{n}/dt$ and the Fano factor $F=\sigma^2/\avg{n}$ can now be
written: \beqn
\label{V}
V&=&\lambda_0^{(1)},\\
\label{F}
F&=&\lambda_0^{(2)}/\lambda_0^{(1)}.  \eeqn
For the MM case (Fig.~\ref{cartoon}A), this gives \beqn
\label{VMM}
V_A&=&V_A^{\rm max}\frac{S}{S+K_A},\\
\label{FMM}
F_A&=&1-\alpha_A\frac{S}{(S+K_A)^2}, \eeqn where $V_A^{\rm max}=k_2$,
$K_A=(k_2+k_{-1})/k_1$, and $\alpha_A=2k_2/k_1$.  The expression for
mean flux $V_A$ is well-known \cite{Michaelis}, and $K_A$ is called
the Michaelis constant; the expression for the Fano factor $F_A$ is
less familiar.

This procedure is fully extendible to other more complicated
enzyme-mediated reactions.  The reaction scheme determines the master
equation and thus $\H^{(0)}$ and $\H^{(1)}$. Specifically, $\hat{H}^{(0)}$
is given by the Markov transition matrix for the enzymatic states
(disregarding the $n$ variable), and $\hat{H}^{(1)}$ has a $1$ marking
every rate where the product gets created, and a $-1$ where it is
destroyed. Then Eqns.\ (\ref{V}-\ref{F}) give the product formation
rate and the Fano factor, and higher orders in perturbation theory
would provide more cumulants. To illustrate the breadth of the method,
in the next section, we apply this procedure to three reaction schemes
that include multiple intermediate reaction steps.

\section{Results: Complex enzymatic reactions}

\subsection{Product distribution statistics}

Many enzyme-mediated reactions involve intermediate steps, and it is
instructive to illustrate our approach with three prototypical examples, shown in Fig.~\ref{cartoon}B-D.

\subsubsection{Reaction scheme B}
 
Fig.~\ref{cartoon}B depicts a case in which the complex undergoes an
intermediate step, such as a conformational change, before creating
the product \cite{frenzen}. This kinetic scheme is also equivalent to
certain ion channels \cite{Bezrukov}.  Such multistep enzymatic
reactions have been shown (including via our method here) to reduce
noise in chemical reactions \cite{Doan}.  The master equation
describing this system is \beqn
\frac{dP^E_n}{dt}&=&-k_1SP^E_n+k_{-1}P^{ES}_n+k_2P^{EP}_{n-1},\\
\frac{dP^{ES}_n}{dt}&=&k_1SP^E_n-(k_{-1}+k_+)P^{ES}_n+k_{-}P^{EP}_n,\\
\frac{dP^{EP}_n}{dt}&=&k_+P^{ES}_n-(k_{-}+k_2)P^{EP}_n, \eeqn which
yields
\ifthenelse{\value{col} = 1}{
\beq \H_B^{(0)} = \begin{pmatrix}
  -k_1 S & k_{-1}	   & k_2			\\
  k_1 S	 & -(k_{-1} + k_{+}) & k_{-}			\\
  0 & k_+ & -(k_{-} + k_2)
\end{pmatrix},
\qquad
\H_B^{(1)} = 
k_2\begin{pmatrix}
0	& 0		& 1	\\
0	& 0		& 0	\\
0	& 0		& 0
\end{pmatrix}.
\eeq
}{
\beqn
 \H_B^{(0)} &=& \begin{pmatrix}
  -k_1 S & k_{-1}	   & k_2			\\
  k_1 S	 & -(k_{-1} + k_{+}) & k_{-}			\\
  0 & k_+ & -(k_{-} + k_2)
\end{pmatrix},\\
\H_B^{(1)} &=& 
k_2\begin{pmatrix}
0	& 0		& 1	\\
0	& 0		& 0	\\
0	& 0		& 0
\end{pmatrix}.
\eeqn
}
The product flux and Fano factor are then
\beqn
\label{VB}
V_B&=&V_B^{\rm max}\frac{S}{S+K_B}\\
\label{FB}
F_B&=&1-\alpha_B\frac{S(S+K'_B)}{(S+K_B)^2}
\eeqn
where $V_B^{\rm max}=k_2k_+/(k_2+k_++k_{-})$, 
$K_B=(k_2k_++k_2k_{-1}+k_{-1}k_{-})/(k_1(k_2+k_++k_{-}))$, 
$\alpha_B=2k_2k_+/(k_2+k_++k_{-})^2$,
and $K'_B=(k_2+k_++k_{-}+k_{-1})/k_1$.

\subsubsection{Reaction scheme C}

Fig.~\ref{cartoon}C depicts a case in which the enzyme exists in an
inactive and an active state. The enzyme switches autonomously between
these states, but can only react with the substrate in its active
form. Note that in this case we have two isolated reactions, since the
enzyme remains in the active state when a product is produced.  This
scheme can be interpreted as a toy model for a voltage-gated ion
channel that can only transmit a single molecule at a time
\cite{hh}. Alternatively, this scheme could be a model for the
production-degradation and subsequent translation of mRNA ($E^*$) by
ribosomes ($S$) into protein ($P$). Finally, this is also an extreme model of an enzyme that has internal states with different rates of product formation, such as studied in \cite{English}. For this scheme we can write the
following master equation: \beqn
\frac{dP^E_n}{dt}&=&-k_+P^E_n + k_{-}P^{E^*}_n\\
\frac{dP^{E^*}_n}{dt}&=&k_+P^E_n-k_{-}P^{E^*}_n+k_2P^{E^*S}_{n-1}
\ifthenelse{\value{col} = 1}{}{\\ &&}
-k_1SP^{E^*}_n+k_{-1}P^{E^*S}_n\\
\frac{dP^{E^*S}_n}{dt}&=&-k_2P^{E^*S}_{n}-k_{-1}P^{E^*S}_n+k_1SP^{E^*}_n
\eeqn which yields
\ifthenelse{\value{col} = 1}{
\beq \H_C^{(0)} =
\begin{pmatrix}
  -k_+ 		& k_{-}			& 0						\\
  k_+ 		& -(k_{-}+k_1 S)	& k_{-1} + k_2			\\
  0		& k_1 S		& -(k_{-1} + k_2)
\end{pmatrix},
\qquad
\H_C^{(1)} = 
k_2\begin{pmatrix}
0	& 0		& 0	\\
0	& 0		& 1	\\
0	& 0		& 0
\end{pmatrix}.
\eeq
}{
\beqn \H_C^{(0)} &=&
\begin{pmatrix}
  -k_+ 		& k_{-}			& 0						\\
  k_+ 		& -(k_{-}+k_1 S)	& k_{-1} + k_2			\\
  0		& k_1 S		& -(k_{-1} + k_2)
\end{pmatrix},\\
\H_C^{(1)} &=& 
k_2\begin{pmatrix}
0	& 0		& 0	\\
0	& 0		& 1	\\
0	& 0		& 0
\end{pmatrix}.
\eeqn
}
The product flux and Fano factor are then
\beqn
\label{VC}
V_C &=& V^{\rm max}_C \frac{S}{S+K_C},\\
\label{FC}
F_C &=& 1 - \alpha_C\frac{S}{(S+K_C)^2}, \eeqn
where $V^{\rm max}_C=k_2$,
$K_C=(k_+{+}k_{-})(k_2+k_{-1})/(k_{+}k_1)$, and
$\alpha_C=2k_2[1+k_-(k_+-k_2-k_{-1})/k_+^2]/k_1$.  Note that these expressions
reduce to those for the MM reaction (Eqns.\ (\ref{VMM}-\ref{FMM})) for
$k_-\to 0$, since this limit corresponds to the enzyme always being in
the active state.  Note also that since $\alpha_C$ can be negative,
$F_C$ can be greater than 1 (and in fact it is infinite in the limit
of rare activation $k_+\to 0$) due to the compounded noise from the
independent stochastic processes of enzyme activation and complex
formation. Under the interpretation of this scheme as protein
translation, $F\gg1$ corresponds to many proteins in a translation
burst from a single rare mRNA.

\subsubsection{Reaction scheme D}

Figure \ref{cartoon}D shows a third example of a more complex reaction
scheme, in which an active enzyme transforms a substrate into a
product and, in contrast to scheme C, returns to its inactive state in
the process.  The enzyme must switch back to its active state for a
new reaction to occur. Similar dynamics have been found for the
$\beta$-galactosidase enzyme \cite{English}. Alternatively, this can be a model for an enzyme that transfers a phosphate group to a
substrate, and needs to reacquire a new phosphate group before
continuing to function as an enzyme.  For this scheme, we can write the
following master equation: \beqn
\frac{dP^E_n}{dt}&=&-k_{+}P^E_n + k_{2}P^{E^*S}_{n-1},\\
\frac{dP^{E^*}_n}{dt}&=&k_{+}P^E_n-k_1SP^{E^*}_n+k_{-1}P^{E^*S}_n,\\
\frac{dP^{E^*S}_n}{dt}&=&k_1SP^{E^*}_n-k_{-1}P^{E^*S}_{n}-k_2P^{E^*S}_n,
\eeqn which yields
\ifthenelse{\value{col} = 1}{
\beqn \H_D^{(0)} =
\begin{pmatrix}
  -k_+	& 0  		& k_2			\\
  k_+		& -k_1S	& k_{-1}		\\
  0		& k_1S	& -(k_{-1} + k_2)
\end{pmatrix},
\qquad
\H_D^{(1)} = 
k_2\begin{pmatrix}
0	& 0		& 1	\\
0	& 0		& 0	\\
0	& 0		& 0
\end{pmatrix}.
\eeqn
}{
\beqn \H_D^{(0)} &=&
\begin{pmatrix}
  -k_+	& 0  		& k_2			\\
  k_+		& -k_1S	& k_{-1}		\\
  0		& k_1S	& -(k_{-1} + k_2)
\end{pmatrix},\\
\H_D^{(1)} &=& 
k_2\begin{pmatrix}
0	& 0		& 1	\\
0	& 0		& 0	\\
0	& 0		& 0
\end{pmatrix}.
\eeqn
}
The product flux and the Fano factor are then
\beqn
\label{VD}
V_D &=& V^{\rm max}_D \frac{S}{S+K_D},\\
\label{FD}
F_D &=& 1 - \alpha_D\frac{S(S+K'_D)}{(S+K_D)^2}, \eeqn where $V^{\rm
  max}_D=k_2k_{+}/(k_2+k_{+})$,
$K_D=k_+(k_2+k_{-1})/(k_1(k_2+k_{+}))$,
$\alpha_D=2k_2k_+/(k_2+k_{+})^2$ and $K'_D=(k_2+k_{+}+k_{-1})/k_1$.
Note that these expressions reduce to those for the MM reaction
(Eqns.\ (\ref{VMM}-\ref{FMM})) for $k_+\to\infty$, since this limit
corresponds to the immediate reversion of the enzyme to its active
state following a product formation.

All four reactions in Fig.~\ref{cartoon} use an enzyme to convert a
substrate into a product, but as we have derived using the present
method, the statistical properties of the product molecule
distributions differ among the cases.

\subsection{Measurable differences between reaction schemes}
Since different reactions have different statistical properties, it
should be possible to use our methods and results to differentiate
among the underlying reactions based on experimental observations.
Here we demonstrate how basic measurements can differentiate among the
four reaction schemes presented above.

\begin{figure}
\centering
\ifthenelse{\value{col} = 1}{
\includegraphics[width = .8\textwidth]{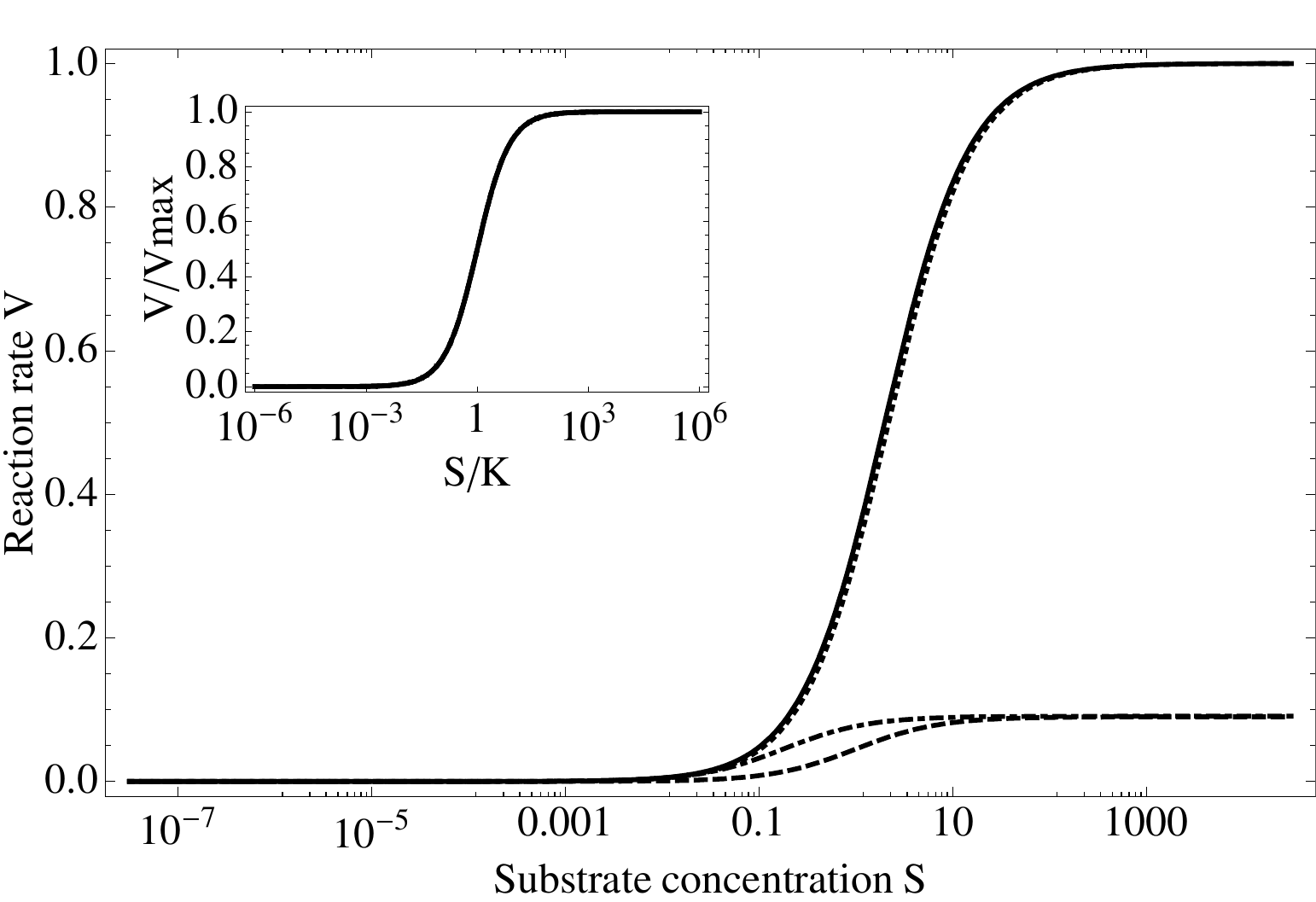}
\includegraphics[width = .8\textwidth]{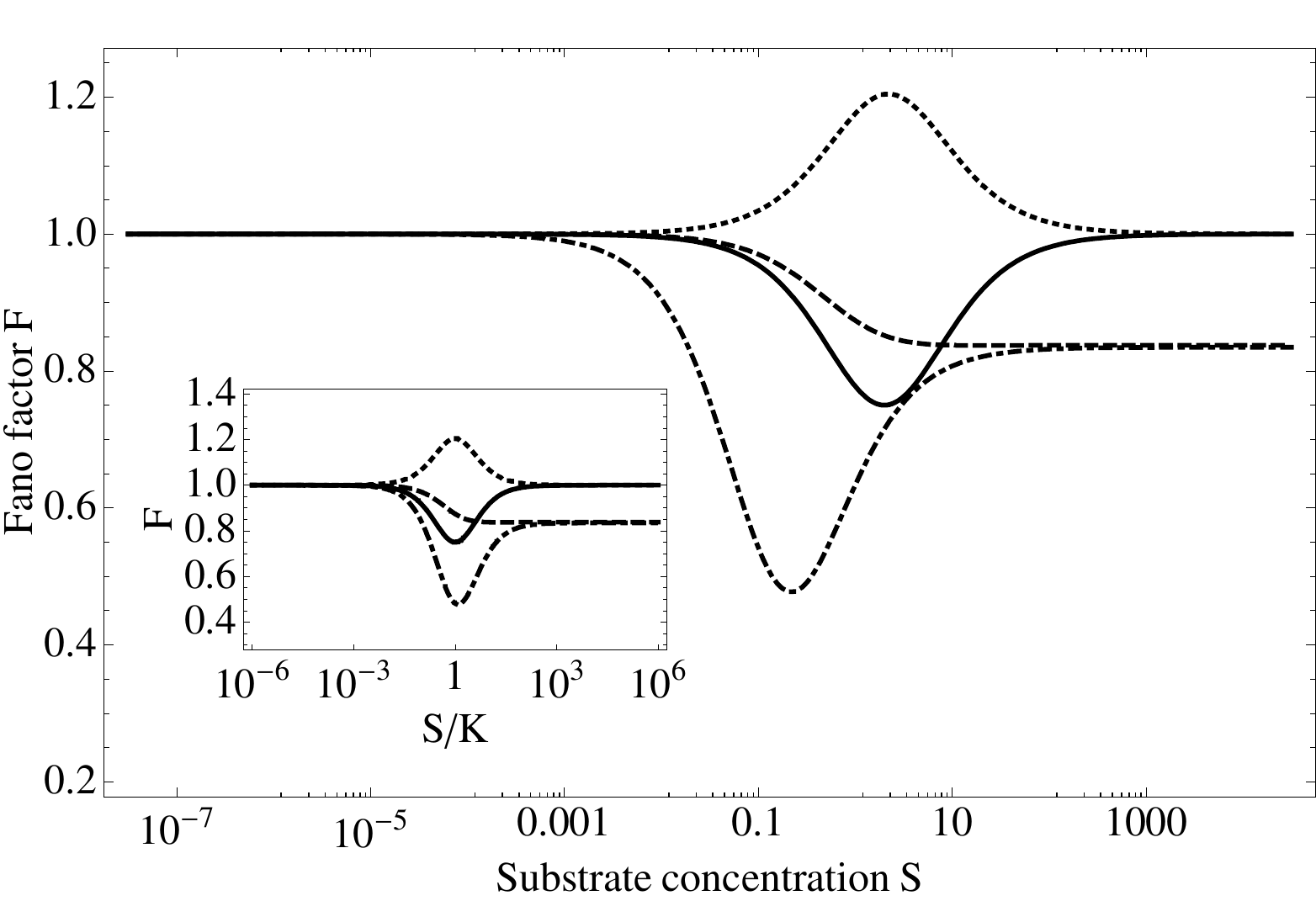}
}{
\includegraphics[width = .47\textwidth]{Vplot_inset_New.pdf}
\includegraphics[width = .47\textwidth]{Fplot_inset_New.pdf}
}
\linespread{1}
\caption{\label{plots} Mean product flux (also called dose-response
  curve) $V$ and the Fano factor $F$ versus substrate concentration
  $S$ for the four cases in Fig.~\ref{cartoon}: A solid, B dashed, C
  dotted, and D dot-dashed.  Plots are of Eqns.\ (\ref{VMM}, \ref{VB},
  \ref{VC}, \ref{VD}) for $V$ and Eqns.\ (\ref{FMM}, \ref{FB},
  \ref{FC}, \ref{FD}) for $F$, with $k_1=1$, $k_{-1}=1$,
  $k_2=1$, $k_+=0.1$, and $k_-=0.01$.  Note that while there are no
  qualitative differences in $V$ (and in fact all curves collapse when
  $V$ is normalized by $V^{\rm max}$ and $S$ by $K$, as seen in the
  inset), features can appear in $F$ that signify that a process is
  more complicated than the single-intermediate case A.}
\end{figure}

The mean product formation rates $V$ for all four reaction schemes A,
B, C and D shown in Fig.~\ref{cartoon}, Eqns.\ (\ref{VMM}, \ref{VB},
\ref{VC}, \ref{VD}), are qualitatively similar functions of substrate
concentration $S$, and it would not be possible to differentiate the
schemes based on mean data alone (see Fig.\ \ref{plots}).  Measurement
of the Fano factor $F$ [Eqns.\ (\ref{FMM}, \ref{FB}, \ref{FC},
\ref{FD})], however, can reveal qualitative and quantitative features
that can differentiate among these schemes, which we outline here and
summarize in Table \ref{tab}.

First, a distinction is possible based on the asymptotic value of $F$
as the substrate concentration $S$ saturates.  For reaction schemes A
and C, \beq F_{A,C}(S\rightarrow\infty)=1, \eeq whereas for reaction
schemes B and D, \beq F_{B,D}(S\rightarrow\infty)=1-\alpha_{B,D}, \eeq
where $\alpha_B$ and $\alpha_D$ are defined following Eqns.\
(\ref{FB}) and (\ref{FD}) respectively.  This expression has a minimum
value $1/2$ in the limits $k_2=k_{+}\gg k_{-}$ for B
and $k_2=k_{+}$ for D.  Thus a saturation value of $F$ that is
significantly less than 1 offers evidence for reaction scheme B or D
over A or C (see Fig.~\ref{plots}).

Second, distinctions are possible based on the value $F^*$ at the
extremum of
the Fano factor as a function of substrate concentration $S$.  For a
MM reaction (case A), there is a minimum:\beq
\label{FstarA}
F^*_A=1-\frac{\alpha_A}{4K_A} = 1-\frac{1}{2}\frac{k_2}{k_2+k_{-1}} ,
\eeq
which is always
between $1/2$ (for $k_2\gg k_{-1}$) and 1 (for $k_{-1}\gg k_2$).
Similarly, for reaction scheme C, we obtain
\beq 
\label{FstarC}
F^*_C=1-\frac{\alpha_C}{4K_C},
\eeq where $\alpha_C$
and $K_C$ are defined following Eqn.\ (\ref{FC}).  This expression
also has a minimum value of $1/2$ (for $k_+\gg k_{-}$ and $k_2 \gg
k_{-1}$), but, unlike in the MM case, it can become larger than 1 if
$k_+(k_++k_-)<k_-(k_2+k_{-1})$ (see Fig.\ \ref{plots}).  Indeed, as
mentioned, in the limit of rare activation $k_+\to 0$, we find
$F^*\rightarrow\infty$.

Depending on the kinetic rates, reaction schemes B and D
may or may not have a minimum for positive $S$ 
(see Fig.\ \ref{plots} for an example of each).
In the cases for which a minimum exists, \beq
\label{FstarB}
F^*_{B,D}=1-\frac{\alpha_{B,D}}{4}\frac{{K'}_{B,D}^2}{K_{B,D}(K'_{B,D}-K_{B,D})},
\eeq where $\alpha_B$, $K_B$, and $K'_B$ are defined following Eqn.\
(\ref{FB}) and $\alpha_D$, $K_D$, and $K'_D$ are defined following
Eqn.\ (\ref{FD}).  This expression has the minimum value $1/3$ in the
limit $k_+=k_2\gg k_{-1}$ for both schemes (and additionally $k_{+}\gg
k_{-}$ for B).  In the reaction scheme B, these limits reduce the
system to a linear irreversible three-step cascade; an $L$-step
irreversible cascade has minimum $F^*$ of $1/L$ in the analogous
limits \cite{Doan}.  Comparing with the MM minimum value of $F^*=1/2$,
it is clear that a measured value of $F^*$ less than $1/2$ is a strong
indication that more than one intermediate step is present
\footnote{In all schemes A, B, C, and D, $F^*$ is dependent on $k_1$
  through $S^*$, which explicitly ensures that $k_1S=k_2$; this is a
  commonly known result, and it may be used by nature to suppress
  noise in natural signaling systems such as phototransduction
  \cite{Doan}.}.

Lastly, distinctions can be made based on measurement of $S^*$, the
substrate concentration at which an extremum in $F$ occurs.
For cases A and C, \beq \frac{S^*_{A,C}}{K_{A,C}}=1, \eeq where $K_A$
and $K_C$ are defined following Eqns.\ (\ref{FMM}) and (\ref{FC})
respectively, and, as in all four cases, $K$ is the concentration at
which $V$ is half-maximal. For cases B and D, on the other hand (when
there is a minimum), \beq
\label{SstarB}
\frac{S^*_{B,D}}{K_{B,D}}=\frac{K'_{B,D}}{K'_{B,D}-2K_{B,D}}, \eeq
where $K_B$ and $K'_B$ are defined following Eqn.\ (\ref{FB}) and
$K_D$ and $K'_D$ are defined following Eqn.\ (\ref{FD}).  This
expression is bounded from below by 1
(e.g.\ for $k_+\gg \{k_-,k_2,k_{-1}\}$ for B, or for $k_+\gg \{k_2,k_{-1}\}$ for D),
but can potentially be infinite
(e.g.\ for $k_-=k_{-1}\gg \{k_2,k_+\}$ for B, or for $k_{-1} \gg k_2=k_+$ for D).
This implies that if
an extremum of the Fano factor occurs at a substrate concentration
significantly different from that at which the mean product formation
rate is half-maximal, it is a strong indication that more than one
intermediate step is present.

Table \ref{tab} summarizes these distinctions, and Figure \ref{plots}
showcases the qualitative differences in the Fano factor curves among
the four reaction schemes caused by differences in the underlying
kinetics. For more complicated reaction schemes, such as multiple
substrate binding by the enzyme, modeled by a high Hill coefficient,
the Fano factor curve would gain even more distinguishing features,
such as additional extrema and/or inflection points
\footnote{We leave this as an exercise for future q-bio Summer School students.}.

\begin{table}
\centering
\begin{tabular}{|c|c|c|c|c|}
\hline
&A		&B		&C	&D				\\ \hline
$F(S\rightarrow\infty)$	&$1$	&$\left[\frac{1}{2}, 1\right]$	&$1$	&$\left[\frac{1}{2}, 1\right]$		\\ \hline
$F^*$		&$\left[\frac{1}{2}, 1\right]$	&$\left[\frac{1}{3}, 1\right]$	&$\left[\frac{1}{2}, \infty\right)$	&$\left[\frac{1}{3}, 1\right]$\\ \hline
$S^*/K$		&$1$		&$\left[1, \infty\right)$	&$1$   &$\left[1, \infty\right)$					\\ \hline
\end{tabular}
\linespread{1}
\caption{Bounds on experimentally measurable quantities that are useful in 
  distinguishing among schemes for enzyme-mediated reactions.  A, B, 
  C, and D refer to reaction schemes in Fig.~\ref{cartoon}.  
  Star ($^*$) denotes the extremum of the Fano factor, such that $F^*$ is the 
  minimum or maximum value and $S^*$ is the substrate concentration at which 
  it occurs.  
  $K$ is the substrate concentration at which product formation rate $V$ is 
  half-maximal.  Generally speaking, minimum bounds on all three quantities 
  occur when forward reaction rates dominate backward rates, and maximum 
  bounds occur when backward rates dominate forward rates; see text for 
  more details.}
\label{tab}
\end{table}

\subsection{Extracting reaction rates from data}
In addition to helping one distinguish among competing reaction
schemes, experimental measurement of the dose-response curve $V(S)$
and the Fano curve $F(S)$ can be used to determine the kinetic rates
of the underlying biochemical reactions.  If the structure of the
biochemical reaction is known, analytical expressions for both curves
in terms of the kinetic rates and the substrate concentration can be
obtained using our method [see e.g.\ Eqns. (\ref{VMM}-\ref{FMM},
\ref{VB}-\ref{FB}, \ref{VC}-\ref{FC}, \ref{VD}-\ref{FD})] and can be
fit to experimental data.  Often times, measurements of the
qualitative features of both curves (such as those highlighted in
Table \ref{tab}) are enough to extract the kinetic rates; for more
complex reactions a full fit to the data would be necessary.
Additionally we note that performing full fits of experimental data to
the analytical expressions may also help in the original task of
distinguishing among (or at least eliminating) different biochemical
reaction schemes.

The MM reaction is an example of a case in which measurement of the qualitative
features is enough to extract all kinetic rates.  However, it is
important to note that in order to do this, one needs both the dose-response
curve and the Fano curve.  In particular, one needs only to measure
the reaction rate at saturation $V^{\rm max}_A$, the substrate
concentration $K_A$ at which the rate is half maximal, and the minimum
value of the Fano curve $F_A^*$.  Then, from Eqn.\ (\ref{FstarA}) and
the expressions following Eqn.\ (\ref{FMM}), one obtains \beqn
k_2&=&V_A^{\rm max},\\
k_{-1}&=&\frac{F_A^*-1/2}{1-F_A^*}V_A^{\rm max},\\
k_1&=&\frac{V_A^{\rm max}}{2K_A(1-F_A^*)}.  \eeqn Instead of obtaining
only $k_2$ and a combination of $k_1$ and $k_{-1}$ by measuring only
the dose-response curve (as is traditionally done for MM reactions),
we now have analytical expressions for all three rates.

For more complex reaction schemes, a similar analysis can be performed
to obtain analytical expressions for the kinetic rates in terms of the
experimental data.  However, it can be the case that not all rates can
be determined unambiguously from measurements of $V$ and $F$ (for the
reaction scheme B, for example, symmetries in the inverted expressions
imply that measurements of $V$ and $F$ do not always uniquely 
determine the five unknown kinetic rates).  
When experimentally feasible, one
may also compare higher moments of the measured product molecule
distributions with those calculated via our method.

\section{Discussion}
We have developed a method of solving chemical master equations for
multistep enzymatic reactions using a perturbation theory approach
analogous to that encountered in quantum mechanics. With this method,
finding cumulants of the distribution of product molecules is
equivalent to diagonalizing a matrix with dimensionality equal to the
number of internal states in the kinetic diagram of the reaction.
Then obtaining the first $m$ cumulants of the reaction can be done by
solving the perturbation theory to $m$th order, which is
straightforward. In particular, the first two moments $\avg{n}$ and
$\sigma^2$ together define the dose-response curve $V=d\avg{n}/dt$ and
the Fano factor $F=\sigma^2/\avg{n}$. As both are currently measurable
in a variety of systems, comparing the calculated $F$ to experimental
data can be used to identify the underlying structure of molecular
reactions.

We have applied this perturbation theory approach to four different
reaction schemes, starting with the simplest Michaelis-Menten
kinetics, and progressing to more complicated kinetic schemes with
internal states. We calculated the dose-response curve and the Fano
factor for each as functions of the substrate
concentration. Importantly, while the dose-response curves for all of
the considered reactions are qualitatively similar, prominent
qualitative features of the Fano factor curve (such as its values at
large substrate concentrations, as well as the position and values
at its extremum) allow us to disambiguate the considered reaction
schemes.  Performing detailed fits of the curves to experimental data
(when feasible) can be an ultimate test for whether the underlying
kinetic structure is known.

For the MM reaction, knowing just a handful of features of the $F(S)$
curve allows us to derive all three rates that completely
define the kinetic scheme, while the entire dose-response curve is
insufficient for this purpose. Similar results hold for the reactions
with intermediate steps, but here the analytical treatment is more
difficult, and often qualitative properties of $F$ alone do not define all of the
underlying kinetic parameters. Instead, a quantitative fit of
derived expressions for $F(S)$ to experimental data would be required.

We stress that the kinetic schemes analyzed in this article are simple
toy models only. However, extending our analysis to more complicated
schemes to derive the first few cumulants of the product number
distribution is not difficult, and it can be automated with just a
simple linear-algebra solver. In particular, calculation of the Fano
factor for a signaling cascade as in \cite{Doan} or for a complex
network of single protein confirmations \cite{Li} is
straightforward. It should be noted, however, that generating
sufficient experimental data to distinguish minute details of
competing kinetic schemes is not easy. Our approach simplifies the
problem somewhat since it does not require single-molecule kinetic
data, as in \cite{English,Li}, but it is based on measuring a
mesoscopic, fluctuating flux. Still, ideally qualitative differences
would dominate the disambiguation task, as emphasized with the toy
models considered here.

\section*{Acknowledgments}
The authors would like to thank the organizers, the lecturers, the
participants, and the sponsors of the $2^{\rm nd}$ q-bio Summer School
on Cellular Information Processing in Los Alamos, NM. We are also
thankful to Michael Wall for careful reading of the manuscript. WdR
was supported by the research program of the ``Stichting voor
Fundamenteel Onderzoek der Materie,'' which is financially supported
by The Netherlands Organization for Scientific Research.  BCD was
supported by NSF Grant DMR-0705167. AM was supported by NSF Grant
DGE-0742450.  NAS and IN were supported by DOE under Contract No.\
DE-AC52-06NA25396. IN was further supported by NSF Grant No.\
ECS-0425850.

\bibliographystyle{apsrev}
\bibliography{mm}

\end{document}